\documentclass[aps,prl,twocolumn,superscriptaddress,showpacs]{revtex4}

\usepackage{times}
\usepackage{mathptmx}
\usepackage{amsmath}
\usepackage{amsfonts}
\usepackage{amssymb}
\usepackage{graphicx}
\usepackage{hyperref}

\newcommand{\ud}{\mathrm{d}}
\newcommand{\beq}{\begin{equation}}
\newcommand{\eeq}{\end{equation}}

\def\addUMD{Maryland Center for Fundamental Physics \& Joint Space-Science Institute, Department of Physics, University of Maryland, College Park, MD 20742}
\def\addRad{Radcliffe Institute for Advanced Study, Harvard University, 8 Garden St., Cambridge, MA 02138}

\begin{document}

\title{Gravitational Self-Force Correction to the Binding Energy of Compact Binary Systems}

\author{Alexandre Le Tiec}
\affiliation{\addUMD}

\author{Enrico Barausse}
\affiliation{\addUMD}

\author{Alessandra Buonanno}
\affiliation{\addUMD}
\affiliation{\addRad}

\begin{abstract}
Using the first law of binary black-hole mechanics, we compute the binding energy $E$ and total angular momentum $J$ 
of two non-spinning compact objects moving on circular orbits with frequency 
$\Omega$, at leading order beyond the test-particle approximation. 
By minimizing $E(\Omega)$ we recover the exact frequency shift of the 
Schwarzschild innermost stable circular orbit induced by the conservative 
piece of the gravitational self-force. Comparing our results for the coordinate 
invariant relation $E(J)$ to those recently obtained from numerical simulations of 
comparable-mass non-spinning black-hole binaries, we find a remarkably good agreement, even in 
the strong-field regime. Our findings confirm that the domain of validity of 
perturbative calculations may extend well beyond the extreme mass-ratio limit.
\end{abstract}

\pacs{04.25.-g,04.25.dg,04.25.Nx,97.60.Lf}

\maketitle

{\it Introduction.---} The problem of motion has always played a central role in physics, especially that of binary systems in gravitational physics. While in Newtonian gravity the two-body orbital motion can be solved analytically, the exact general relativistic solution for two black holes was only obtained very recently, by means of numerical simulations~\cite{NR}.

The post-Newtonian (PN) approximation to General Relativity has proven a
valuable tool to describe the dynamics of widely separated
inspiralling compact-object binaries~\cite{Bl.06}, which are expected to be the main
gravitational-wave (GW) sources for existing ground-based interferometric detectors
(LIGO/Virgo) and future space-based antennas. Furthermore, it has
achieved successes ranging from
Solar-system tests of gravity~\cite{Wi.06}, to measurements of
the damping rate of binary pulsars due to GW emission~\cite{TaWe.89}, and to remarkable practical applications such as
the Global Positioning System (GPS)~\cite{As.03}. However, the PN approximation 
becomes inaccurate during the late stages of the binary's inspiral,
and breaks down during the final plunge and merger.
While numerical-relativity (NR) simulations can describe these highly relativistic phases of the evolution,
they are still too time-consuming to construct GW template 
banks covering the whole parameter space of compact binaries.

An alternative analytical approach that can improve our knowledge of 
this highly relativistic regime is the gravitational self-force (GSF) formalism, a natural extension of black-hole perturbation theory \cite{Po.al.11}. The GSF approach relies on an expansion in the binary's mass ratio, and is the natural tool to model extreme mass-ratio compact binaries, which are among the most promising GW sources for space-based detectors~\cite{Am.al.07}. Unlike the PN approximation, which breaks down when the binary's velocity becomes close to the speed of light, the GSF formalism remains valid for highly relativistic systems, even in the strong-field regime. Moreover, it has recently become clear that the GSF can also describe at least certain aspects of the dynamics of \textit{comparable-mass} binary systems, such as the relativistic periastron advance \cite{Le.al.11}, possibly allowing the construction of vaster and more accurate template banks that would be crucial for GW astronomy.

In this Letter, we confirm that picture by computing the binding energy $E$ and angular momentum $J$ of a non-spinning circular-orbit compact binary system within the GSF formalism. Our results provide a surprisingly accurate description of comparable-mass binaries, as we verified by comparing the coordinate-invariant relation $E(J)$ that we obtain with recent NR data \cite{Da.al.11}. In addition, they pave the way to adiabatic evolutions of extreme mass-ratio inspirals that include the effect of the first-order conservative GSF.

More specifically, we start from the first law of mechanics for binaries of spinless compact objects moving along circular orbits, and modelled as point particles. This relation, recently established in Ref.~\cite{Le.al2.11}, gives the variations of the total Arnowitt-Deser-Misner (ADM) mass $M$ and total angular momentum $J$ of the binary system in response to small variations of the individual masses $m_A$ ($A = 1,2$) of the compact objects according to
\beq\label{first_law}
	\delta M - \Omega \, \delta J = z_1 \, \delta m_1 + z_2 \, \delta m_2 \, .
\eeq
Here we set $G=c=1$ (a choice that we adopt throughout this Letter), $\Omega$ is the circular-orbit frequency, and $z_A$ are 
the so-called ``redshift observables'', namely the gravitational redshifts of light rays emitted from the particles, 
and received far away from the binary along the direction perpendicular to the orbital plane \cite{De.08}. 
In a convenient gauge, the redshifts simply coincide with the inverse time components of the four-velocities $u_A^\alpha$ of the particles, 
namely $z_A = 1 / u_A^t$ \cite{De.08}. The relation \eqref{first_law} is the point-particle analog 
of the celebrated first law of black-hole mechanics 
$\delta M - \Omega_{\rm H} \, \delta J = 4 \kappa\, m_\text{irr} \, \delta m_\text{irr}$ \cite{Ba.al.73}, 
where $m_\text{irr} = \sqrt{A / (16\pi)}$ is the irreducible (or Christodoulou) mass 
of a black hole of mass $M$, spin $J = a\, M$, surface area $A$, 
uniform surface gravity $\kappa$, and horizon frequency $\Omega_{\rm H}$. 
(See Refs.~\cite{Fr.al.02,Le.al2.11} for more details on the first law of black-hole mechanics, or ``thermodynamics''.)

In this Letter we show that the first law \eqref{first_law} can be used, 
in conjunction with existing perturbative GSF calculations of the redshift observable, 
to compute the binding energy $E \equiv M - (m_1 + m_2)$ and angular momentum $J$ of compact binary systems on circular orbits, 
at leading-order beyond the test-particle approximation. 
(See Ref.~\cite{Le.al2.11} for a discussion of the applicability of Eq.~\eqref{first_law} to GSF calculations.) 

As an immediate application of our results, we 
recover the \textit{exact} frequency shift of the Schwarzschild 
innermost stable circular orbit (ISCO) induced by the conservative piece of the GSF, 
as computed previously from a stability analysis of slightly eccentric orbits around a non-spinning black 
hole~\cite{BaSa.09}. We then compare our newly derived GSF-accurate expression for the invariant relation $E(J)$ 
with the results of spinless binary black-hole simulations with mass ratios $q\equiv m_1 / m_2 = 1$, $1/2$, and $1/3$ (we assume $m_1 \leqslant m_2$ throughout this Letter). Finally, we summarize our results and discuss future work.

{\it Binding energy and angular momentum.}---The ADM mass $M$, total angular momentum $J$, and redshift observables $z_A$ are 
all functions of the circular-orbit frequency $\Omega$ and the individual masses $m_A$. 
The first law \eqref{first_law} thus implies $\partial M / \partial \Omega = \Omega \, \partial J / \partial \Omega$ 
and $\partial M / \partial m_A - \Omega \, \partial J / \partial m_A = z_A$. Applying the change of variables $(\Omega, m_1, m_2) \to (m, \nu, x)$, where $m \equiv m_1 + m_2$ is the total mass, $\nu \equiv m_1 m_2 / m^2 \equiv \mu / m$ the symmetric mass ratio, and $x \equiv (m\, \Omega)^{2/3}$ the usual dimensionless invariant PN parameter, these relations can be combined to give
\begin{subequations}\label{z1_E_J}
	\begin{align}
		m \, z_1 &= \mathcal{M} + \frac{2x}{3} \frac{\partial \mathcal{M}}{\partial x} + \frac{1-4\nu+\Delta}{2} \, \frac{\partial \mathcal{M}}{\partial \nu} \, , \label{z1} \\
		M &= \mathcal{M} - \frac{2x}{3} \frac{\partial \mathcal{M}}{\partial x} \, , \!\!\quad J = - \frac{2m}{3\sqrt{x}} \frac{\partial \mathcal{M}}{\partial x} \, ,
	\end{align}
\end{subequations}
where $\mathcal{M} \equiv M - \Omega\, J$ can heuristically be viewed as the energy of the binary in 
a co-rotating frame, and $\Delta \equiv (m_2 - m_1) / m = \sqrt{1-4\nu}$ is the reduced mass difference.
To derive Eq.~\eqref{z1}, we used the fact that the dimensionless ratio $\mathcal{M}/m$ is a function of $\nu$ and $x$ only (\textit{cf.} the discussion following Eq.~(4.21) of Ref.~\cite{Le.al2.11}).
A similar equation for particle $2$ can be obtained by substituting $\Delta \to - \Delta$.

At first order beyond the test-particle approximation, the following mass-ratio expansions hold for the redshift $z_1$,
the specific binding energy $\hat{E} \equiv (M - m) / \mu$, the dimensionless angular momentum $\hat{J} \equiv J / (m \mu)$,
and $\hat{\mathcal{M}} \equiv (\mathcal{M} - m) / \mu$:
\begin{subequations}\label{exp}
	\begin{align}
		z_1 &= \sqrt{1-3x} + \nu \, z_{\rm SF}(x) + {\cal O}(\nu^2) \, , \label{z1exp} \\
		\hat{E} &= \biggl( \frac{1-2x}{\sqrt{1-3x}} - 1 \biggr) + \nu \, E_{\rm SF}(x) + {\cal O}(\nu^2) \, , \label{Eexp} \\
		\hat{J} &= \frac{1}{\sqrt{x(1-3x)}} + \nu \, J_{\rm SF}(x) + {\cal O}(\nu^2) \, , \label{Jexp} \\
		\hat{\mathcal{M}} &= (\sqrt{1-3x} - 1) + \nu \, \mathcal{M}_{\rm SF}(x) + {\cal O}(\nu^2) \, . \label{Mexp}
	\end{align}
\end{subequations}
The lowest-order terms are the well-known Schwarzschild results in the test-mass limit. Since $\nu = q / (1+q)^2 = q+{\cal O}(q)^2$, these equations
remain valid if the symmetric mass ratio $\nu$ is replaced by the usual mass ratio $q$.
Substituting the expansions~\eqref{exp} in Eqs.~\eqref{z1_E_J}, one obtains the following relations between the various GSF corrections:
$z_{\rm SF}(x) = 2 \mathcal{M}_{\rm SF}(x) - 2(\sqrt{1-3x}-1) - x / \sqrt{1-3x}$, 
$E_{\rm SF}(x) = \mathcal{M}_{\rm SF}(x) - \frac{2x}{3} \mathcal{M}'_{\rm SF}(x)$, 
and $J_{\rm SF}(x) = - \frac{2}{3\sqrt{x}} \mathcal{M}'_{\rm SF}(x)$, 
where we denote $\mathcal{M}'_{\rm SF} \equiv \ud \mathcal{M}_{\rm SF} / \ud x$. Eliminating $\mathcal{M}_{\rm SF}$, we then find
\begin{subequations}\label{SF}
	\begin{align}
		E_{\rm SF}(x) &= \frac{1}{2} \, z_{\rm SF}(x) - \frac{x}{3} \, z'_{\rm SF}(x) - 1 \nonumber \\ &\qquad\quad\;\; + \sqrt{1-3x} + \frac{x}{6} \frac{7-24x}{(1-3x)^{3/2}} \, , \label{E_SF} \\
		J_{\rm SF}(x) &= - \frac{1}{3\sqrt{x}} \, z'_{\rm SF}(x) + \frac{1}{6\sqrt{x}} \frac{4-15x}{(1-3x)^{3/2}} \, . \label{J_SF}
	\end{align}
\end{subequations}
The knowledge of the GSF correction $z_{\rm SF}$ to the redshift 
(and thus of its first derivative $z'_{\rm SF} = \ud z_{\rm SF} / \ud x$) 
therefore immediately gives the corrections $E_{\rm SF}$ and $J_{\rm SF}$ to the test-particle results for the binding energy and angular momentum.

The GSF effect on the coordinate-invariant relation $z_1(x)$ was first computed
in the Regge-Wheeler gauge in Ref.~\cite{De.08}. Alternative GSF calculations based on 
different gauges (harmonic gauge and radiative gauge) 
were later found in agreement to within the numerical uncertainties \cite{Sa.al.08,Sh.al.11}. 
Collecting all the published GSF data~\cite{De.08,Sa.al.08,Sh.al.11,Bl.al.10}, 
we have at our disposal $55$ data points for  $z_{\rm SF}(x)$, with $x$ ranging from $0$ to $1/5$,
and with relative errors lower than $10^{-6}$.
 In particular, $14$ of these points lie in the ``strong-field interval'' 
$5m \leqslant r \leqslant 10m$, where $r \equiv m / x$ is an invariant measure of the separation. 

Following Ref.~\cite{Ba.al.10}, it is convenient to represent these data by a compact analytic expression, 
in terms of a ratio of polynomials in the PN parameter $x$. We adopt the model 
 $z_{\rm SF}(x) = 2 x \, (1 + a_1 x + a_2 x^2) / (1 + a_3 x + a_4 x^2 + a_5 x^3)$, 
which accounts for the asymptotic behavior $z_{\rm SF}(x) = 2 x + \mathcal{O}(x^2)$ when 
$x \to 0$ \cite{Le.al2.11}. Performing a standard least-squares fit, we find that the 
coefficients $a_1 = -2.18522$, $a_2 = 1.05185$, $a_3 = -2.43395$, $a_4 = 0.400665$, and $a_5 = -5.9991$ reproduce
the data to within $10^{-5}$. We notice that the extrapolation of our 
fit beyond the data point with the smallest separation
($r = 5m$) diverges very close to $r = 3m$. This 
suggests that the GSF contribution to $z_1$ may have a pole near the Schwarzschild 
circular photon-orbit (or ``light-ring''), located at $x_{\rm LR} = 1 / 3$.

Using this fit for $z_{\rm SF}(x)$ in Eqs.~\eqref{SF}, the GSF contributions $E_{\rm SF}$ and $J_{\rm SF}$ to the binding energy $\hat{E}$ and angular momentum $\hat{J}$ can easily be computed in the range $0 \leqslant x \leqslant 1/5$, with a comparable accuracy.

{\it Self-force correction to the Schwarzschild ISCO.---} 
We recall that the Schwarzschild ISCO is defined by the onset of a dynamical (radial) instability for circular orbits. 
As a first application of our results, we show how the GSF-induced shift of the ISCO frequency
can be recovered very simply from the expressions \eqref{Eexp} and \eqref{E_SF} for $\hat{E}(\Omega)$. 

Beyond the test-particle approximation, the orbital frequency of the ISCO is given by 
\beq
	m\,\Omega_{\rm ISCO} = 6^{-3/2} \left[ 1 + \nu \, C_\Omega + {\cal O}(\nu^2) \right] ,
\eeq
where the coefficient $C_\Omega$, which encodes the effect of the conservative piece of the GSF,
has recently been computed by Barack and Sago (BS) \cite{BaSa.09,BaSa.10}. 
Performing a stability analysis of slightly eccentric orbits near $r = 6m$, 
they found $C^{\rm BS}_\Omega = 1.2512(4)$ \cite{BaSa.10}. 
This strong-field benchmark has since then been used as a
reference point for comparison with other analytical and numerical methods \cite{Lo.al.10,Fa.11},
and for calibrations of the effective-one-body (EOB) model~\cite{Da.10,Ba.al.10,BaBu.10,Pa.al.11}.

On the other hand, the minimum-energy circular orbit (MECO) is defined as the minimum of the binding energy $\hat{E}(\Omega)$; 
the MECO's orbital frequency $\Omega_{\rm MECO}$ thus satisfies
\beq\label{MECO}
	\frac{\partial \hat{E}}{\partial \Omega} \Big|_{\Omega_{\rm MECO}} = 0 \, .
\eeq
It was shown in Ref.~\cite{Bu.al.03} that the notions of ISCO and MECO are formally equivalent; hence $\Omega_{\rm ISCO} = \Omega_{\rm MECO}$.
This result does not rely on any PN expansion or perturbative analysis, and thus holds for any mass ratio, even in the strong-field regime;
it only requires that the binary's dynamics can be derived from a Hamiltonian.
Inserting Eqs.~\eqref{Eexp} and \eqref{E_SF} for the binding energy $\hat{E}(\Omega)$ in the condition \eqref{MECO}, 
we then find the following expression for the ISCO frequency shift:
\beq\label{C_Omega}
	C_\Omega = \frac{1}{2} + \frac{1}{4 \sqrt{2}} \left[ \frac{1}{3} \, z''_{\rm SF}\bigg(\frac{1}{6}\bigg) - z'_{\rm SF}\bigg(\frac{1}{6}\bigg) \right] .
\eeq
In order to compute the GSF correction to the Schwarzschild ISCO, 
one thus only needs the first and second derivatives of the GSF correction to the redshift, evaluated at $x = 1/6$. 
By fitting our full data set for $z_{\rm SF}(x)$ (or the strong-field subset $5m \leqslant r \leqslant 10m$ only) 
 to different models, we find $C_\Omega = 1.2510(2)$, which agrees with BS's result $C^{\rm BS}_\Omega$ at the $1\sigma$ level.
 
Moreover, Eq.~\eqref{C_Omega} allows for a \textit{highly} accurate determination of the ISCO frequency shift. 
Indeed, while previous calculations relied on a stability analysis near a singular point, 
using a GSF code capable of handling slightly eccentric orbits, Eq.~\eqref{C_Omega} only requires the 
evaluation of a \textit{regular} function near $x = 1/6$, using a much simpler GSF code 
for \textit{circular} orbits. Current circular-orbit GSF codes implemented in the frequency domain can 
already deliver highly accurate results: for instance, the data reported in Refs.~\cite{Bl.al.10,Bl.al2.10} 
for the GSF correction to $u_1^t = 1 / z_1$ (for separations $r \geqslant 200m$) are accurate to 
within $10^{-13}$, and similar accuracies should be achievable at least down to $r=5m$. With high-accuracy GSF data for $z_{\rm SF}$ near $x = 1/6$, 
it will become possible to determine the ISCO frequency shift $C_\Omega$ much more accurately than ever before; a valuable result given the physical significance of this genuinely strong-field effect. In particular, this could prove useful to cross-check the results of different GSF codes.

\begin{figure}
\includegraphics[scale=0.4]{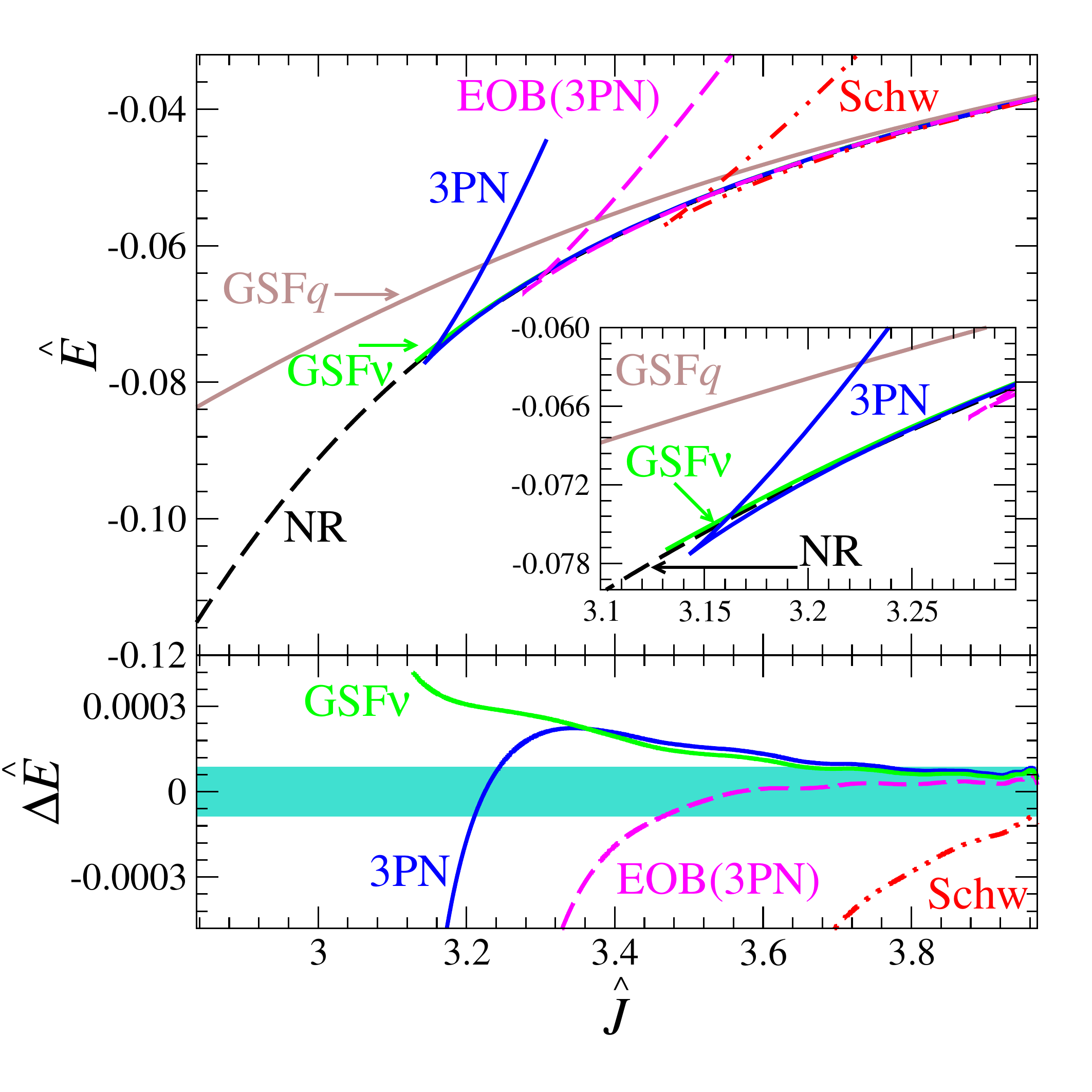}
\caption{\footnotesize In the upper panel, the (specific) binding energy $\hat{E} = E / \mu$ of an equal-mass black-hole binary is shown as a function of the (dimensionless) total angular momentum $\hat{J} = J / (\mu m)$, as computed in numerical relativity (``NR''), in PN theory (``3PN''), in the EOB model [``EOB(3PN)''], in the test-particle approximation (``Schw''), and including the conservative gravitational self-force (``GSF$q$'' and ``GSF$\nu$''). The 3PN, EOB(3PN), and test-particle curves show cusps at their respective ISCO/MECO; the lower branches correspond to stable circular orbits, while the upper branches correspond to unstable circular orbits. We might find similar branches for the GSF curves when more data closer to the light-ring become available. The ``GSF$q$'' and ``GSF$\nu$'' curves are only shown in the region where numerical data for the self-force  are available (\textit{i.e.} $r \geqslant 5m$, corresponding to $\hat{J} \simeq 3.126$ for the ``GSF$\nu$'' model and to $\hat{J} \simeq 1.896$ for the ``GSF$q$'' model). The differences between the various models and the NR result, $\Delta \hat{E} \equiv \hat{E} - \hat{E}_{\rm NR}$, are shown in the lower panel (down to the ISCO/MECO, when that is present), with the exception of the ``GSF$q$'' model, which quickly grows beyond the plot range. The shaded area represents the error affecting the NR results.}
\label{fig:EJ}
\end{figure}

{\it Comparison with numerical relativity.---}We now combine our results \eqref{exp} and \eqref{SF} for the binding energy 
$\hat{E}(\Omega)$ and angular momentum $\hat{J}(\Omega)$ to compute the coordinate-invariant relation $\hat{E}(\hat{J})$, 
at leading-order beyond the test-mass approximation, and compare it to the results recently obtained in Ref.~\cite{Da.al.11}
using accurate NR simulations \cite{Po.al3.11} (with Cauchy characteristic extraction \cite{Re.al.09}) of non-spinning black-hole binaries with mass ratios $q = 1, 1/2$, and $1/3$.
This comparison is presented in Fig.~\ref{fig:EJ} for the $q=1$ case, and gives similar results for $q=1/2$ and $q=1/3$.

The curve labelled ``Schw'' shows the  relation $\hat{E}(\hat{J})$ for a test mass on an adiabatic sequence of circular orbits around a Schwarzschild black hole, given in parametric form by the lowest-order terms in Eqs.~\eqref{Eexp} and \eqref{Jexp}. It necessarily presents a cusp at the Schwarzschild ISCO, located at $m\,\Omega^{\rm Schw}_{\rm ISCO} \simeq 0.068$. As can be seen in Fig.~\ref{fig:EJ}, this result is in reasonable agreement with the NR data all the way  down to the ISCO, where the difference reaches $\Delta \hat{E} \simeq -1.7 \times 10^{-3}$. The astonishing agreement between the  test-mass result and the NR result for $q=1$ suggests that the relation $\hat{E}(\hat{J})$ should be almost independent of $q$, at least for ``large'' orbital separations. Indeed, we verified this using the NR data for $q=1$, $1/2$, and $1/3$, as well as the other models shown in Fig.~\ref{fig:EJ}.

The $\hat{E}(\hat{J})$ adiabatic relation given in parametric form by
Eqs.~\eqref{Eexp} and \eqref{Jexp} includes the effect of the
conservative GSF. Expressing Eqs.~\eqref{Eexp} and \eqref{Jexp} in terms
of the mass ratio $q$ (which simply amounts to replacing $\nu \to q$), we obtain the
curve ``GSF$q$''.  Given the large mass ratio involved
($q=1$), the poor agreement with NR is expected, and the agreement
does not improve significantly for $q=1/2$ or $q=1/3$.  However, the
GSF result expressed in terms of the \textit{symmetric} mass ratio
$\nu$ (``GSF$\nu$'') compares remarkably well with
the NR result, with a difference that grows larger than
the numerical error only near $r
= 5m$ (which corresponds to $\hat{J} \simeq 3.126$ for the ``GSF$\nu$'' model, where it reaches 
$\Delta \hat{E} \simeq 3.5\times 10^{-4}$).

For completeness, we also show the invariant relation $\hat{E}(\hat{J})$ as computed in the adiabatic PN approximation
(``3PN'') and in the EOB adiabatic model [``EOB(3PN)''].
The PN result is given in the parametric form
$\{\hat{E}(\Omega),\hat{J}(\Omega)\}$ by, \textit{e.g.},
  Eqs.~(5.13) and (5.8) of Ref.~\cite{Da.al.00}, where we set $\nu =
  1/4$, and use the known values $\omega_\text{static} = 0$ and
  $\omega_\text{kinetic} = 41/24$ for the 3PN static and kinetic
  ``ambiguity parameters''.  We observe very good agreement with the
NR data all the way down to the cusp occurring at the 3PN MECO
($m\,\Omega^{\rm 3PN}_{\rm MECO} \simeq 0.129$), where the difference
grows to $\Delta \hat{E} \simeq -10^{-3}$.  
In that respect, we point out that this PN approximant performs much better
than the PN approximant considered in Ref.~\cite{Da.al.11} [see Eq.~(5) there],
which is obtained through the series 
reversion $\hat{E}[\Omega(\hat{J})]$ and subsequent re-expansion through 
3PN order. Although technically correct, the resulting (uni-valued) function 
necessarily fails to capture the cusp at the MECO. Moreover, we checked 
that even in the test-particle limit the parametric form 
$\{\hat{E}(\Omega),\hat{J}(\Omega)\}$ is closer than the expression
$\hat{E}[\Omega(\hat{J})]$ to the Schwarzschild result.  The poor 
behavior of some PN approximants, such as $\hat{E}[\Omega(\hat{J})]$, 
is not a surprise in PN theory (see, \textit{e.g.}, Ref.~\cite{boyle_et_al}).
The EOB result in Fig.~\ref{fig:EJ} is produced using the 3PN model of
Ref.~\cite{Da.al3.00}, with a (1,3) Pad{\'{e}} model for the effective
metric component $g^\text{eff}_{tt}$ (see, \textit{e.g.},
Ref.~\cite{Da.al.11}).  As can be seen the difference with respect to
the NR data grows as large as $\Delta \hat{E} \simeq - 9 \times
  10^{-4}$ near the cusp occurring at the EOB ISCO
($m\,\Omega^{\rm EOB}_{\rm ISCO} \simeq 0.088$).

We emphasize that the NR curve was obtained by Ref.~\cite{Da.al.11}
from an actual binary black-hole evolution, and therefore includes
non-adiabatic effects during the late inspiral and plunge. These effects are not captured 
by our adiabatic models, and may in part explain the 
differences from the NR result in Fig.~\ref{fig:EJ} at small 
$\hat{J}$ (\textit{i.e.} at large $\Omega$). We also note that the remarkable agreement between 
the adiabatic models and the NR data for $\hat{E}(\hat{J})$
does not automatically imply that the same will hold true for the
invariant relations $\hat{E}(\Omega)$ and $\hat{J}(\Omega)$.

Finally, we stress that although the GSF-accurate binding energy necessarily has 
an ISCO/MECO for small  mass ratios (since for $q=0$ it reduces
to the binding energy of a test mass in a Schwarzschild background), 
it does not present an ISCO/MECO (at least for $r \geqslant 5m$)  
for $q = 1$, $1/2$, and $1/3$. It remains to be seen if this 
holds true even when GSF data for $r<5m$ become available.

{\it Summary and future work.---}Recently, the general relativistic periastron advance of non-spinning black-hole binaries on quasi-circular orbits was computed in NR, and compared to the prediction of the GSF (as well as to other approximation techniques)~\cite{Le.al.11}. By expressing the GSF result in terms of the symmetric mass ratio $\nu$ rather than the usual mass ratio $q$, the GSF prediction was found in remarkable agreement with the exact NR result, even for comparable masses. This prompted the authors of Ref.~\cite{Le.al.11} to suggest that the domain of validity of perturbative calculations may extend well beyond the extreme mass-ratio limit. Our new, alternative comparison based on the invariant relation $\hat{E}(\hat{J})$ strongly supports this expectation. A similar observation was previously made for the dissipative component of the GSF, based on a comparison of perturbative and NR calculations of the GW energy flux for head-on collisions \cite{Det_Sma}. The ``scaling-up'' procedure $q \to \nu$ has also been used in the context of perturbative calculations of the linear momentum flux for quasi-circular orbits \cite{FiDe.84,Fa.al.04}. In the future, our analysis should be revisited using more GSF data in the very strong-field region $3m < r \leqslant 5m$, as well as including dissipative effects in a consistent GSF evolution \cite{Wa.al.11}.

Our expression for the binding energy $\hat{E}(\Omega)$ can also be used to compute 
the EOB effective metric component $g^\text{eff}_{tt}$ exactly, through linear order in the mass ratio.
Furthermore, by combining this result with the 
recent GSF/EOB comparison of Ref.~\cite{Ba.al.10} for the periastron advance in quasi-circular compact binaries,
the $g^\text{eff}_{rr}$ component of the EOB effective metric can also be computed. These results, which
completely determine the EOB metric for spinless binaries through linear order in the mass ratio, are presented in the companion paper \cite{Ba.al.11}.

In the Schwarzschild spacetime, circular orbits for massive particles exist for any radius $r > 3m$. 
The redshift observable can thus be calculated, at least in principle, at any such radius. When more data 
for $z_{\rm SF}(x)$ near $x = 1/3$ become available, our formulas~\eqref{exp} and \eqref{SF} for 
the binding energy and angular momentum will provide information about the shift of the light-ring 
frequency induced by the conservative GSF acting on ultra-relativistic particles, or photons.

In this respect, we emphasize that the connections established by Eqs.~\eqref{SF} are particularly useful 
to explore the highly relativistic regime: while standard perturbative analyses cannot describe the binary's 
dynamics beyond the ISCO \cite{BaSa.11}, the relations \eqref{SF} give \textit{direct} access to 
the binding energy $E$ and angular momentum $J$ in the very strong-field regime $3m < r \leqslant 6m$, 
using only ``routine'' GSF calculations of the redshift $z_1$ for circular orbits.

\vspace{0.1cm}

We wish to thank T. Damour, A. Nagar, D. Pollney, and C. Reisswig for sharing with us the numerical relativity data used to produce Fig.~\ref{fig:EJ}. All three authors acknowledge support from NSF Grant No. PHY-0903631. A.B. also acknowledges support from NASA Grant No. NNX09AI81G, and A.L.T. from the Maryland Center for Fundamental Physics.

\bibliography{}

\end{document}